# OXYGEN REDUCTION ACTIVITY OF CARBON NITRIDE SUPPORTED ON CARBON NANOTUBES

S. M. Lyth, Y. Nabae, N. M. Islam, S. Kuroki, M. Kakimoto and S. Miyata

*Department of Organic and Polymeric Sciences, Tokyo Institute of Technology, Ookayama, Meguro-ku, Tokyo 152-8550, Japan. Email: lyth@physics.org*

Fuel cells offer an alternative to burning fossil fuels, but use platinum as a catalyst which is expensive and scarce. Cheap, alternative catalysts could enable fuel cells to become serious contenders in the green energy sector. One promising class of catalyst for electrochemical oxygen reduction is iron-containing, nanostructured, nitrogen-doped carbon. The catalytic activity of such N-doped carbons has improved vastly over the years bringing industrial applications ever closer. Stoichiometric carbon nitride powder has only been observed in recent years. It has nitrogen content up to 57% and as such is an extremely interesting material to work with. The electrochemical activity of carbon nitride has already been explored, confirming that iron is not a necessary ingredient for 4-electron oxygen reduction. Here, we synthesize carbon nitride on a carbon nanotube support and subject it to high temperature treatment in an effort to increase the surface area and conductivity. The results lend insight into the mechanism of oxygen reduction and show the potential for carbon nanotube-supported carbon nitride to be used as a catalyst to replace platinum in fuel cells.

**INTRODUCTION**

Fuel cells are increasingly being used in real-life applications.[1] Recently, polymer electrolyte membrane fuel cells (PEMFCs) utilizing the oxygen reduction reaction have received attention as clean, green power sources, due to their simplicity, feasibility, and quick start-up.[2] The price and scarcity of platinum has driven-up the cost of PEMFCs.[3] Therefore, research in alternative non-precious catalysts is gathering momentum.[4]

One contender in this field is nitrogen-containing carbon, often prepared via pyrolysis of Fe- or Co-containing $N_4$-macrocycles.[5] These carbon-based catalysts are getting ever closer to replacing platinum in commercial applications.[6] Such materials are a complex mixture of carbon, nitrogen, and iron. As such, the nature of the active site for 4e⁻ oxygen reduction is still much debated, mainly as to whether the active site involves $FeN_2C$ and/or $FeN_4C$ sites [6-9],[7-10] or that some carbon-nitrogen atomic configuration is active,[11-21] the role of Fe being to stabilize the incorporation of nitrogen into the carbon lattice.

Carbon nitride powder[22-24] (Figure 1) could be an ideal precursor for such materials, due to its extremely high nitrogen content. Using carbon nitride as an oxygen reduction catalyst has already lent insight into the nature of the active site, suggesting that iron is not necessary for 4-electron oxygen reduction.[25] Here we attempt to increase the electrochemical oxygen reduction activity by supporting the carbon nitride on multiwall carbon nanotubes and pyrolysing at elevated temperature. Multiwall carbon nanotubes are well-known for their high surface area and conductivity, making them ideally suited as catalyst-supports.

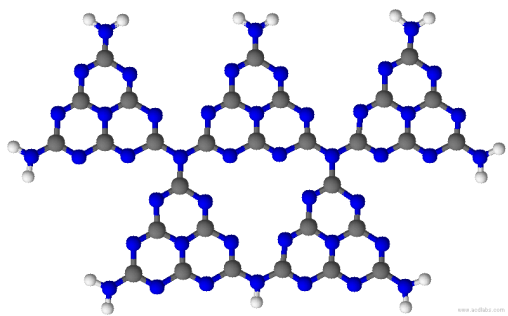

Figure 1 [Colour]. Structure of graphitic carbon nitride.

**EXPERIMENTAL**

Carbon nitride was synthesized in a magnetically stirred, stainless-steel high pressure reactor.[26] As-received multiwall carbon nanotubes (*Cheaptubes.com*) were also added to the reaction chamber as a high surface-area catalyst support. Subsequently the material was milled and then subject to pyrolysis at a range of temperatures in $N_2$. The electrochemical activity was measured via rotating ring-disk electrode voltammetry at room temperature in a 0.5 M $H_2SO_4$ electrolyte solution.[26] The *onset potential* is defined here at -2 μA/cm$^2$.

The nitrogen content was measured using CHN elemental analysis to be 6.93, 2.77, 0.97 and 0.39 wt% in the as-produced material and after pyrolysis at 600, 800 and 1000°C, respectively. CHN underestimates the nitrogen content of the carbon nitride at the surface, due to the underlying carbon nanotube support which is made up almost entirely of carbon. CHN analysis of separate, pure, carbon nitride produced in the same manner typically yields a nitrogen content of over 50 wt%.[26]

X-ray Photoelectron Spectroscopy (XPS) was measured to give insight into the type of bonding in the catalysts. The N1s signal (Fig. 2) probably comprises contributions from pyridinic (~398.5 eV), pyrrolic (~399.5 eV), quaternary (~400.5 eV) and amine (~401.4 eV) – type nitrogen bonds. It appears from the N1s signal that the proportion of pyridinic nitrogen decreases as the temperature increases, whilst the proportion of pyrrolic and/or quaternary nitrogen increases. Meaningful quantitative deconvolution is difficult for such a small signal.

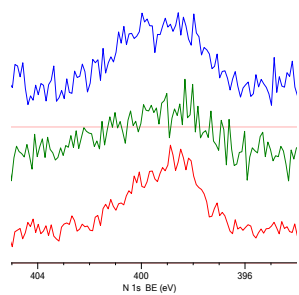

Figure 2 [Colour]. X-ray Photoelectron Spectroscopy (XPS) N1s signal of carbon nitride samples pyrolysed at 600°C (red), 800°C (green), and 1000°C (blue).

A Transmission Electron Microscope (TEM) image of the carbon black-supported carbon nitride sample pyrolysed at 1000°C is shown in Figure 3. Because the two materials are largely carbon, it is difficult to observe any contrast change between carbon nitride and carbon black. The nanotubes appear to be slightly buckled, and are shorter than the as-purchased samples, because of a combination of the magnetic stirring in the high-pressure reactor, and the subsequent milling process.

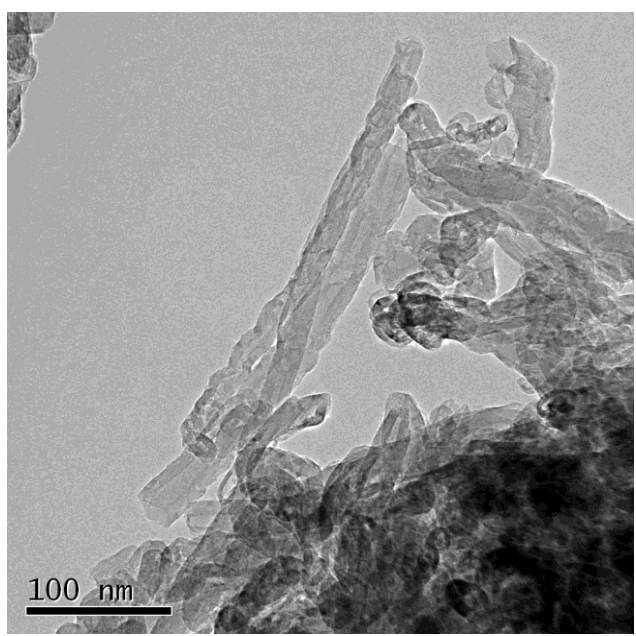

Figure 3. Transmission Electron Microscopy (TEM) image of carbon nitride supported on multiwall carbon nanotubes and pyrolysed at 1000°C.

The electrochemical behavior of the pyrolysed nanotube-supported carbon nitride samples is presented in Figure 4, along with data for multiwall carbon nanotubes alone, and unsupported carbon nitride pyrolysed at 1000°C. There is strong correlation between the pyrolysis temperature and the current density, which increases dramatically between the as-produced sample (red line) and that pyrolysed at 800°C (blue line). This increase in current density may be associated with an increase in surface area and/or conductivity as the carbon nitride decomposes. After pyrolysis at 1000°C (pink line) there is a slight decrease in the oxygen reduction current density, probably due to either a reduction in surface area, or the falling nitrogen content resulting in a reduction in the number of available active sites. The brown line shows the oxygen reduction activity of un-supported carbon nitride pyrolysed at 1000˚C. It is evident that by supporting carbon nitride on multiwall carbon nanotubes, the oxygen reduction current density is vastly improved, likely due to the increase in surface area, and an increase in electrical conductivity. All carbon nitride samples perform better than multiwall carbon nanotubes alone (black line), showing that the carbon nitride is a vital component in the catalytic activity of these samples. However, for applications the current density needs to be improved, and this is the subject of further work.

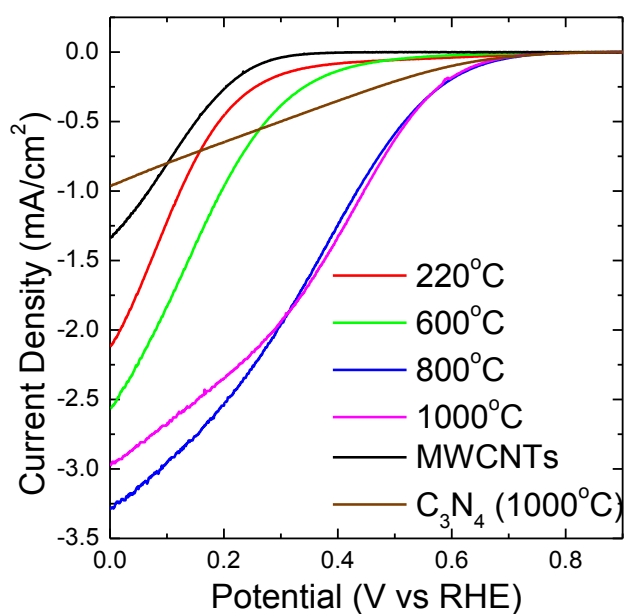

Figure 4 [Colour]. Linear-sweep oxygen-reduction voltammogram of pyrolysed carbon nitride catalysts.

The onset potential gives more information about the inherent catalytic activity, but no trend is observed with heating, with all the samples displaying an onset potential in the range 0.80 to 0.85 V. The best onset potential is 0.85 V is high for this class of catalyst and indicates a 4-electron pathway (2-electron oxygen reduction would result in a maximum onset potential of 0.695 V). The onset potential of un-supported carbon nitride is also 0.85 V, suggesting that the catalytic activity is due to the carbon nitride coating, rather than the underlying multiwall carbon nanotubes. The onset potential of multiwall carbon nanotubes alone is just 0.43 V, further confirming that the nanotubes themselves are not catalytically active. Taking into account the information gleaned about bonding using XPS, it is possible that the relative enrichment of quaternary and/or pyrrolic nitrogen is responsible for the enhanced catalytic activity. Recent results using hard X-ray photoemission spectroscopy (HXPES) suggest that quaternary-type nitrogen is indeed important for electrochemical oxygen reduction.[27]

The ring potential was maintained at 1.1 V throughout electrochemical measurements, in order to oxidize any hydrogen peroxide produced via the 2-electron pathway, resulting in a measurable current associated with $H_2O_2$ production. From this, the number of electrons ($n$) transferred can be estimated according to the formula; $n = 4I_{disk} / (I_{disk} + I_{ring}/N)$, where $I_{disk}$ and $I_{ring}$ are the currents at the disk and ring electrodes, and N is the estimated efficiency of hydrogen peroxide oxidation.[28] Here, the average number of electrons transferred per event is 3.3 for our best sample. This suggests that oxygen reduction proceeds via a mix of the 2-electron and 4-electron pathways. For applications, the value should be nearer to 4-electron transfer and this will be the subject of further work with these catalysts.

Cyclic voltammograms for samples are shown in Figure 5. There is a pair of redox peaks in the data for all samples, probably due to oxygen-containing functionalities,[29] which have no affect on oxygen reduction.[27] It is evident from the anodic shift in onset potential (where the $O_2$ curve deviates significantly from the $N_2$ curve) that the electrochemical properties are affected by pyrolysis. There is also an increase in the double-layer capacitance of the voltammogram up to 800˚C, suggesting an increase in physical surface area of the catalyst. This corresponds well with the linear sweep voltammogram in which the current density increased with increasing pyrolysis temperature. This double-layer capacitance decreases slightly after pyrolysis at 1000°C, suggesting a slight decrease in surface area and also agreeing with the linear sweep voltammogram in Figure 4.

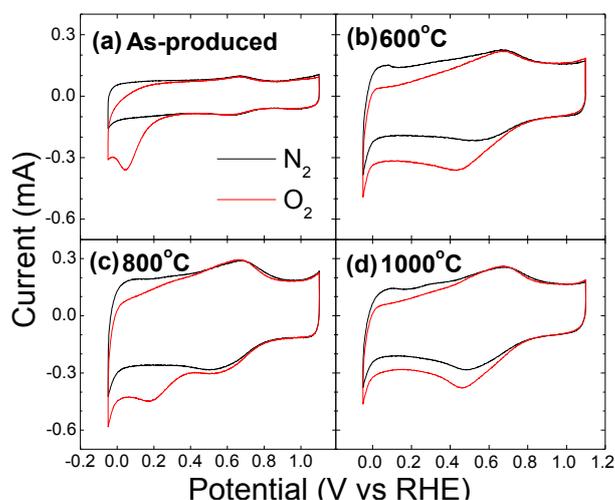

Figure 5 [Colour]. Cyclic voltammograms of; (a) as-prepared samples, and those pyrolysed at (b) 600°C, (c) 800°C, and (d) 1000°C.

In an attempt to further increase the performance of these catalysts, a variation on the pyrolysis process was attempted, which has been successful with our other catalysts.[6] Firstly, the samples were pyrolysed at 600°C for a longer time (5 hours) and subject to a further milling step. The blue line in Figure 6 shows that this greatly improves the current density when compared with the sample pyrolysed at 600°C for just 1 hour (black line). This may be due to an increase in crystallization and reorientation of bonds associated with the longer pyrolysis step, or due to an increase in surface area associated with the milling step. After this treatment, the sample was subject to a further heat treatment in flowing $NH_3$ gas (as opposed to nitrogen) at 800°C for 1 hour. This has a detrimental effect on the current density of the sample, perhaps due to a change in surface area resulting from etching by the ammonia gas. However, the onset potential is increased from 0.8 V to 0.89 V, which can be seen in the enlarged data (inset). This increase in onset potential with ammonia treatment has been observed previously.[6] By further heat treatment steps at higher temperature it is expected that the surface area, and therefore the current density, can be increased. This is the subject of further research.

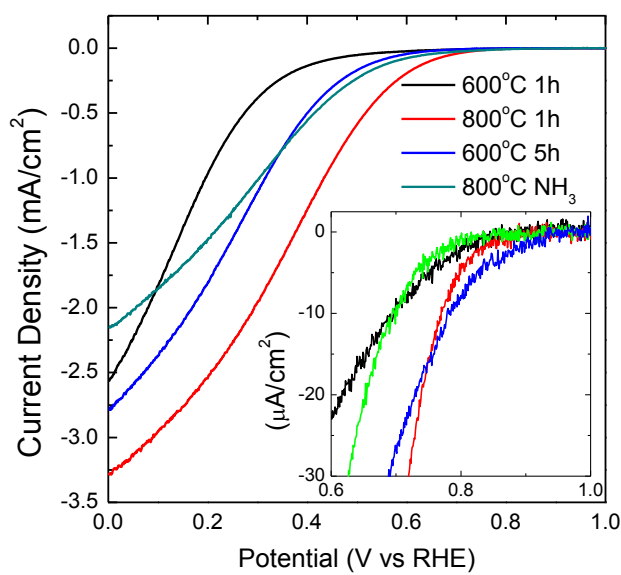

Figure 6 [Colour]. Linear sweep voltammograms of carbon nitride subjected to secondary and tertiary pyrolysis steps.

## SUMMARY

In summary, carbon nitride supported on multiwall carbon nanotubes, was pyrolyzed at up to 1000°C. The initial nitrogen content was 6.93 at.%, falling to 0.39 at.% after pyrolysis at 1000°C, measured by CHN. The proportion of pyridinic nitrogen was found to decrease after pyrolysis, whilst the proportion of quaternary and/or pyrrolic nitrogen increased. The optimum treatment temperature for electrochemical activity was found to be 800°C. The onset potential for oxygen reduction was 0.85 V, and the average number of electrons was 3.3, suggesting that oxygen reduction occurs mainly via a 4e$^-$ pathway. The reason for this high activity is attributed to enriched quaternary nitrogen at this temperature. This work shows that carbon nitride supported on multiwall carbon nanotubes has excellent potential for use in PEMFCs, after further research and optimization.

**AKNOWLEDGEMENTS**

The authors thank the New Energy and Industrial Technology Development Organization (NEDO) for financial support, the Center for Advanced Materials Analysis (Tokyo Institute of Technology) for characterization. We also thank Yo Hosaka, Mayu Sonoda and Chiharu Yamauchi for technical assistance.